\documentclass[12pt,a4paper]{article}
\usepackage{epsfig}
\begin{document}
\title{Single production of charged gauge bosons from little\\
\hspace*{-0.8cm} Higgs models in association with top quark at the
$LHC$}

\author{Chong-Xing Yue, Shuo Yang, Li-Hong Wang \\
{\small  Department of Physics, Liaoning Normal University, Dalian
116029, China}\thanks{E-mail:cxyue@lnnu.edu.cn}\\}
\date{\today}

\maketitle
\begin{abstract}

In the context of the little Higgs models, we discuss single
production of the new charged gauge bosons in association with top
quark at the $CERN$ Large Hadron Collider$(LHC)$. We find that the
new charged gauge bosons $W_{H}^{-}$ and $X^{-}$, which are
predicted by the littlest Higgs model and the $SU(3)$ simple model,
respectively, can be abundantly produced at the $LHC$. However,
since the main backgrounds coming from the processes $pp\rightarrow
t\bar{t}+X$ and $pp\rightarrow tW^{-}+X$ are very large, the values
of the ratios $N_{W}$ and $N_{X}$ are very small in most of the
parameter space.  It is only possible to detect the signal of the
gauge boson $W_{H}^{-}$ via the process $pp\rightarrow
gb+X\rightarrow tW_{H}^{-}+X$ at the $LHC$ in a small region of the
parameter space.

\end{abstract}

\vspace{2.0cm} \noindent
 {\bf PACS number(s)}:  12.60.Cn, 14.65.Ha, 12.15.-y

\newpage
\noindent{\bf I. Introduction}

 Most of the new physics models
predict the existence of the new gauge bosons, generally called $Z'$
or $W'$ with masses in the $TeV$ range. If these new particles are
discovered, they would represent irrefutable proof of new physics,
most likely that the gauge group of the standard model ($SM$) must
be extended. Thus, search for extra gauge bosons provides a common
tool in quest for new physics at the next generation collider
experiments[1]. The discovery and study of extra gauge bosons is one
of the important goals of the $CERN$ Large Hadron Collider ($LHC$).

The little Higgs models[2] propose a new approach to electroweak
symmetry breaking ($EWSB$) accomplished by a naturally light Higgs
sector. These models predict new gauge bosons, fermions, and scalars
at or below the $TeV$ scale, which might generate characteristic
signatures at the next generation collider experiments, especially
at the $LHC$[3,4]. The little Higgs models generally predict the
existence of the pure left-handed charged gauge boson $W'$, which
has the $SM$-like couplings to ordinary particles and might generate
significant corrections to single top quark production at the
$LHC$[5]. It has been shown that the charged gauge bosons from
little Higgs models should be either seen or excluded in the first
year of running at the $LHC$[6]. Thus, it is very interesting to
study production of these new charged gauge bosons at the $LHC $.

In the $SM$ framework, single production of the charged gauge bosons
$W^{\pm}$ associated with a top quark is one of the important
process for single top quark production. Its cross section is
negligible at the $Tevatron$ but of considerable size at the $LHC$,
where it is larger than that of the $s$-channel process for single
top quark production. There is an extensive work on $tW$ associated
production at the $LHC$ in the literature[7,8,9]. The production
cross section for this process has been calculated at leading order,
with a subset of the next-to-leading order ($NLO$) corrections
included[7]. Including both the subsequent leptonic decays
$W\rightarrow l\nu$ and $t\rightarrow l\nu b$, as well as the
emission of real radiation in the top decay, a $NLO$ calculation of
this process has been given in Ref.[8]. Recently, Ref.[9] gives a
complete calculation of electroweak supersymmetric effects at one
loop to $tW$ associated production at the $LHC$. In this letter, we
will study single production of the charged gauge bosons predicted
by the little Higgs models in association with top quark and see
whether the possible signals of these new particles can be detected
via this process at the $LHC$.

\noindent{\bf II. The relevant formulas}

 Based on the structure of the extended electroweak gauge group, the little Higgs models can be
generally divided into two classes[3,10]: the product group models
in which the $SU(2)_{L}$ gauge group arises from the diagonal
breaking of two or more gauge groups, and the simple gauge group
models in which the $SU(2)_{L}$ gauge group arises from the breaking
of a single larger gauge group down to an $SU(2)$ subgroup. The
littlest Higgs models ($LH$)[11] and the $SU(3)$ simple group
model[10,12] are the simplest examples of the product group models
and the simple group models, respectively. All of these little Higgs
models predict the pure left-handed charged gauge boson $W'$. The
couplings of the new particle $W'$ to fermions can be unitive
written as:

\begin{equation}
\pounds_{W'q_{i}q_{j}}=\frac{eA}{\sqrt{2}S_{W}}V_{ij}W'_{\mu}\bar{q_{i}}r^{\mu}P_{L}q_{j}
 ,
\end{equation}
where $P_{L}=(1-\gamma_{5})/2$, $V_{ij}$ is the $CKM$ matrix
element, and $S_{W}=\sin\theta_{W}$ ($\theta_{W}$ is the Weinberg
angle). For the charged gauge boson $W^{-}_{H}$ predicted by the
$LH$ model, the constant $A$ equals to $c/s$ for
$(q_{i},q_{j})=(u,d),(c,s)$, and $(t,b)$, in which
$c(s=\sqrt{1-c^{2}})$ is the mixing parameter between $SU(2)_{1}$
and $SU(2)_{2}$ gauge bosons[13]. For the charged gauge boson
$X^{-}$ predicted by the $SU(3)$ simple group model, the constant
$A$ equals to $\delta_{t}$ and $\delta_{\nu}$ for
$(q_{i},q_{j})=(t,b)$ and $(u,d)$[or $(c,s)$], respectively. The
factors $\delta_{t}$ and $\delta_{\nu}$ can be written as[3]:
\begin{eqnarray}
\delta_{t}=\frac{\nu}{\sqrt{2}f}t_{\beta}\frac{x_{\lambda}^{2}-1}{x_{\lambda}^{2}+t_{\beta}^{2}},\
\ \ \delta_{\nu}=-\frac{\nu}{2ft_{\beta}}.
\end{eqnarray}
Where $\nu=246GeV,\ f=\sqrt{f_{1}^{2}+f_{2}^{2}}$,
$t_{\beta}=\tan\beta=f_{2}/f_{1}$, and
$x_{\lambda}=\lambda_{1}/\lambda_{2}$, in which $f_{1}$ and $f_{2}$
are the vacuum condensate values of the two sigma-model fields
$\Phi_{1}$ and $\Phi_{2}$, respectively. $\lambda_{1}$ and
$\lambda_{2}$ are the $Yukawa$ coupling parameters.

It has been shown that the collinear $\bar{b}$ component of the
inclusive process $gg\rightarrow t \bar{b} W^{-}$ can be attached as
a $QCD$ $NLO$ correction to bottom quark distribution function of
the exclusive process $bg\rightarrow t  W^{-}$ at the partonic
level[7,8,9]. The $LO$ contributions to $tW^{-}$ associated
production is best considered to arise from the process
$bg\rightarrow t  W^{-}$. Thus, in this paper, we only consider
single production of the heavy charged gauge boson $W^{-}_{i}$ in
association with a top quark via the two body final state process
$bg\rightarrow t  W^{-}_{i}$ at the $LHC$ as shown in Fig.1 in which
$i=1$ and 2 represent the gauge bosons $W_{H}^{-}$ and $X^{-}$,
respectively. Our numerical results are easy transferred to those
for single production of the heavy gauge boson $W^{+}_{i}$ at the
$LHC$ by replacing $\bar{b}$ as $b$ and $\bar{t}$ as $t$.

\begin{figure}[htb]
\vspace{-6cm}\hspace{-5cm}
\epsfig{file=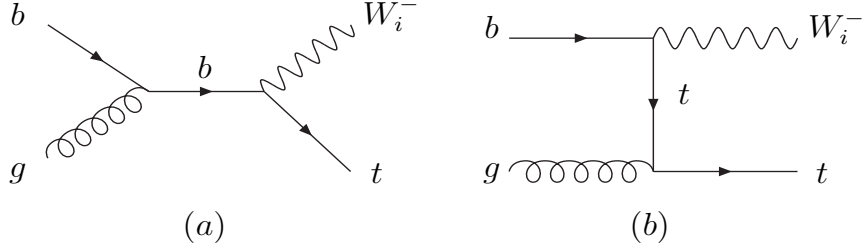,width=650pt,height=850pt} \vspace{-21cm}
\caption{Feynman diagrams for the partonic process $bg\rightarrow t
W^{-}_{i}$}
\end{figure}

At the leading order, the scattering amplitude of the partonic
process $bg\rightarrow t  W^{-}_{i}$ can be written as:
\begin{equation}
M=\frac{eg_{s}A}{\sqrt{2}S_{W}}\bar{u}(t)[\frac{\not\varepsilon_{2}
P_{L} (\not P_{g}+\not P_{b}+m_{b}) \not\varepsilon_{1}} {\hat
s'-m_{b}^{2}}+\frac{\not\varepsilon_{1}(\not P_{t}-\not
P_{g}+m_{t})\varepsilon_{2}P_{L}} {\hat u-m_{t}^{2}}]u(b),
\end{equation}
where $\hat s'=(P_{g}+P_{b})^{2}=(P_{W_{i}}+P_{t})^{2}$, $\hat
u=(P_{t}-P_{g})^{2}=(P_{b}-P_{W_{i}})^{2}$. $\varepsilon_{1}$ and
$\varepsilon_{2}$ are the gluon and $W^{-}_{i}$ polarization
vectors, respectively.

The cross section for single production of the heavy gauge boson
$W_{i}^{-}=W_{H}^{-}$ or $X^{-}$ associated with a top quark at the
$LHC$ with the center-of-mass $\sqrt{S}=14TeV$ can be obtained by
convoluting the production cross section $\hat \sigma(tW^{-}_{i})$
of the partonic process $gb\rightarrow t  W_{i}^{-}$ with the parton
distribution functions ($PDF's$):

\begin{equation}
\sigma_{i}(S)=\int^{1}_{\tau}dx_{1}\int^{1}_{\tau/x_{1}}dx_{2}
[f_{g/p}(x_{1},\mu) f_{b/p}(x_{2},\mu) \hat\sigma(t W_{i}^{-})+
f_{b/p}(x_{1},\mu) f_{g/p}(x_{2},\mu) \hat\sigma(t W_{i}^{-})],
\end{equation}
where $\tau=(M_{W_{i}}^{2}+m_{t}^{2})/S$ and $\hat s'= x_{1}x_{2}S$
. Through out this paper, we will use $CTEQ6L$ $PDF's$[14] for the
bottom quark and gluon $PDF's$. Following the suggestions given by
Refs.[7,8,9], we assume that the factorization scale $\mu$ for the
bottom quark $PDF$ is of order $(M_{W_{i}}+m_{t})/4$.

\noindent{\bf III. Numerical results and conclusions}

 Except for the
$SM$ input parameter $m_{t}=171.4GeV$[15], $\alpha_{e}=1/128.8$,
$\alpha_{s}=0.118$, and $S_{W}^{2}=0.2315$[16], the production cross
sections of the new charged gauge bosons $W_{H}^{-}$ and $X^{-}$
coming from the $LH$ model and the $SU(3)$ simple group model are
dependent on the free parameters ($M_{W_{H}}$, $c$) and ($M_{X}$,
$x_{\lambda}$, $t_{\beta}$), respectively. Considering the
constraints of the electroweak precision data on these free
parameters, we will assume $1TeV \leq M_{W_{H}} \leq 2TeV$ and $0 <
c \leq 0.6 $ for the $LH$ model[17] and $1TeV \leq M_{X} \leq 2TeV$,
$x_{\lambda} >1$, and $t_{\beta}
>1$ for the $SU(3)$ simple group model[3,10,12] in our numerical
calculation.

\begin{figure}[htb] \vspace{-0.5cm}
\begin{center}
\epsfig{file=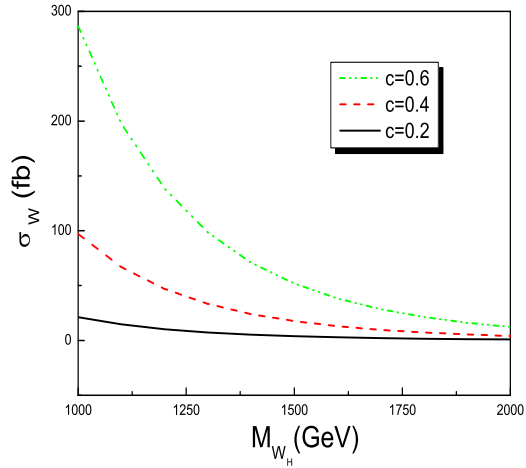,width=220pt,height=205pt}
\hspace{-0.5cm}\vspace{-1cm}
 \caption{The cross section $\sigma_{W}$ as a function of the mass parameter $ M_{W_{H}} $ for the mixing
 \hspace*{1.8cm}parameter $ c=$0.2(solid line), 0.4(dashed line), and 0.6(dashed-dotted line).  }
 \label{ee}
\end{center}
\end{figure}

The cross sections $\sigma_{W}$ and $\sigma_{X}$ for single
production of the new gauge bosons $W_{H}^{-}$ and $X^{-}$
associated with top quark at the $LHC$ with $\sqrt{S}=14TeV$ are
plotted as functions of the mass parameters $M_{W_{H}}$ and $M_{X}$
in Fig.2 and Fig.3, respectively, in which we have taken different
values of the free parameters $c $, $x_{\lambda}$, and $ t_{\beta}$.
One can see from these figures that the single production cross
section for the gauge boson $W_{H}^{-}$ is larger than that for the
gauge boson $X^{-}$, which is because compared to the coupling
$W_{H}^{-}t\bar{b}$, the coupling $X^{-}t\bar{b}$ is suppressed by
the factor $\nu/f$. For $0.2\leq c\leq 0.6$, $1TeV\leq M_{W_{H}}\leq
2TeV$, the value of the production cross section $\sigma_{W}$ is in
the range of $0.7fb\sim 286.7fb$. The production cross section
$\sigma_{X}$ increase as the free parameter $x_{\lambda}$
increasing. For $x_{\lambda}=5$, $1\leq t_{\beta}\leq 3$, and
$1TeV\leq M_{X}\leq 2TeV$, the value of $\sigma_{X}$ is in the range
of $0.16fb\sim15.3fb$.

\begin{figure}[htb] \vspace{-0.5cm}
\begin{center}
\epsfig{file=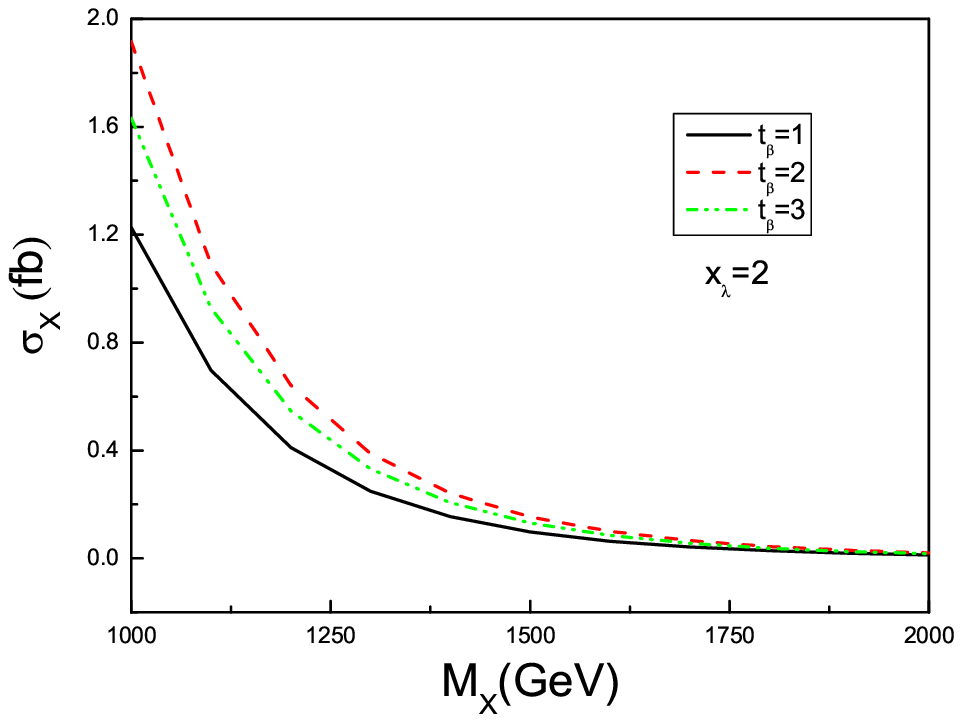,width=220pt,height=200pt} \put(-110,3){
(a)}\put(115,3){ (b)}
 \hspace{0cm}\vspace{-0.25cm}
\epsfig{file=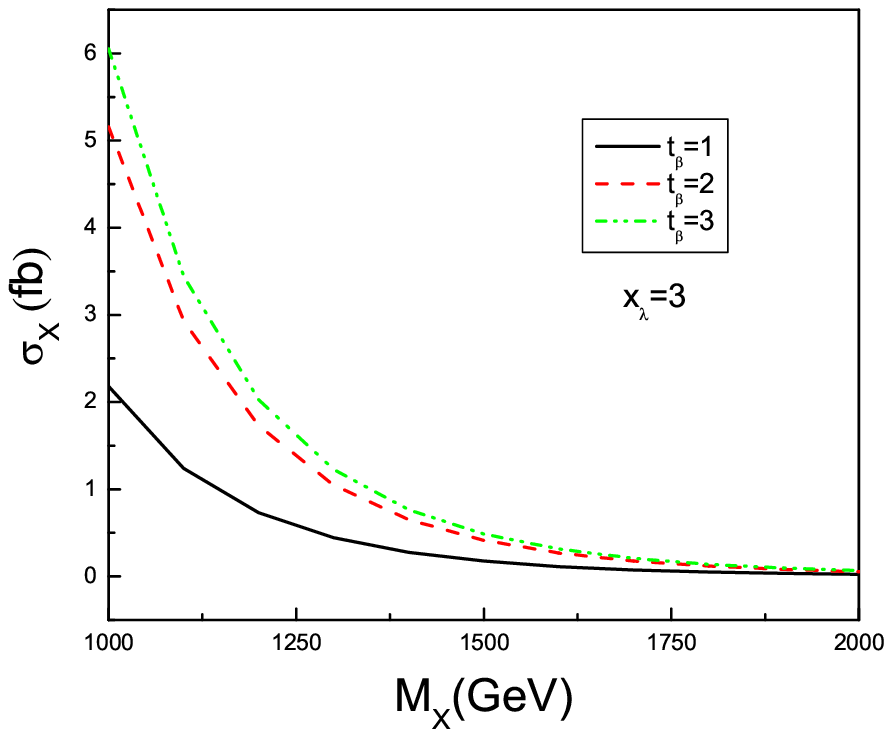,width=220pt,height=200pt} \hspace{-0.5cm}
 \hspace{10cm}\vspace{-1cm}
\epsfig{file=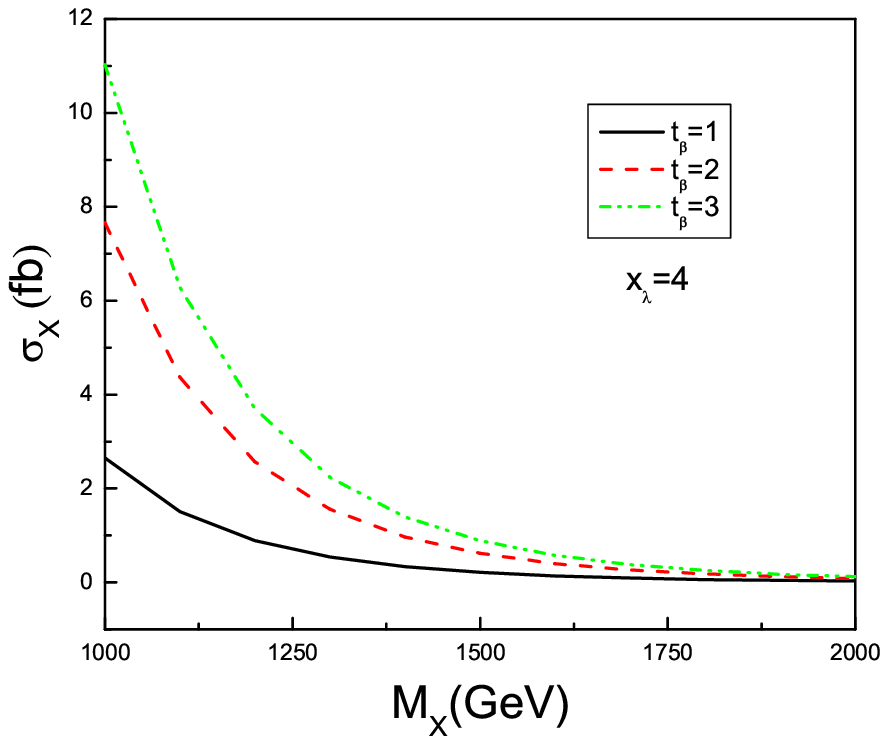,width=220pt,height=200pt} \hspace{-0.5cm}
 \hspace{0cm}\vspace{-0.25cm}
 \put(-110,3){(c)}\put(115,3){ (d)}
\epsfig{file=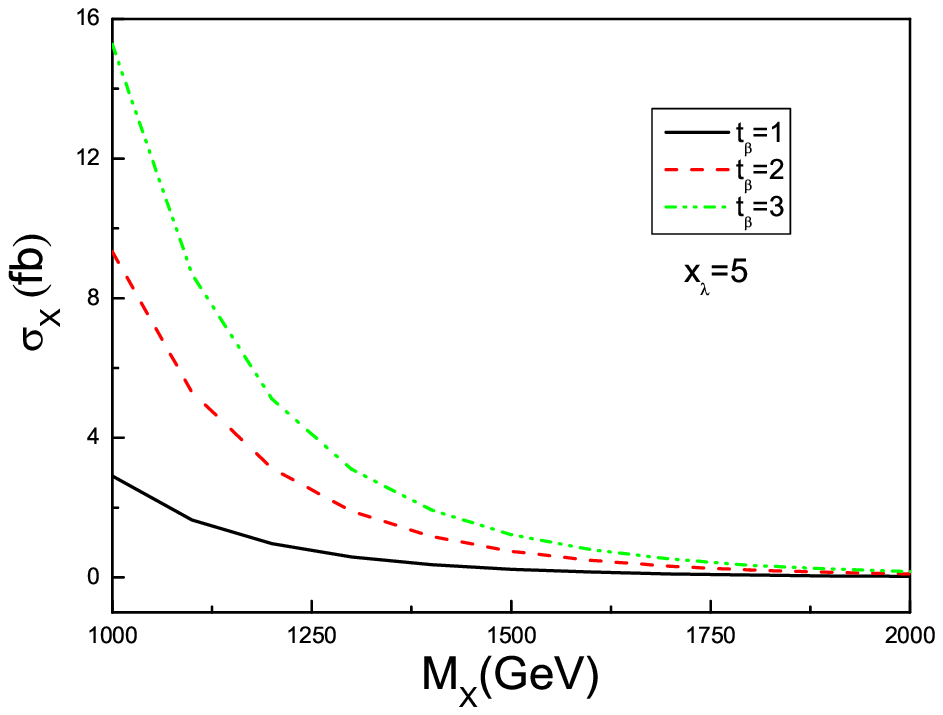,width=220pt,height=200pt} \hspace{-0.5cm}
 \hspace{10cm}\vspace{-0.2cm}
 \caption{The cross section $\sigma_{X}$ as a function of the mass parameter $ M_{X} $ for
 different \hspace*{1.8cm} values of the free parameters  $x_{\lambda}$ and $ t_{\beta}$.}
 \label{ee}
\end{center}
\end{figure}
\vspace{-0.5cm}

To see whether the heavy gauge bosons $W_{H}^{-}$ and $X^{-}$ can be
observed at the $LHC$ via the processes $pp\rightarrow gb
+X\rightarrow t W_{H}^{-} +X$ and $pp\rightarrow gb +X\rightarrow
tX^{-}+X$, we discuss the possible decay modes of the heavy gauge
bosons $W_{H}^{-}$ and $X^{-}$. For $W_{H}^{-}$, the main decay
modes are $W^{-}H$, $\bar{q}q'$, and $l^{-}\nu$, in which $q(q')$
and $l^{-}$ present all three generation quarks and leptons,
respectively. If we neglect the final state masses, then we have
$\Gamma(W_{H}^{-}\rightarrow \bar{q}q')=3\Gamma(l^{-}\nu)$. At
leading order, the total decay width $\Gamma _{W_{H}^{-}}$ can be
approximately written as[17]:
\begin{equation}
\Gamma
_{W_{H}^{-}}=\frac{\alpha_{e}}{96S_{W}^{2}}[\frac{96c^{2}}{s^{2}}+\frac{C_{W}^{2}(c^{2}-s^{2})^{2}}{s^{2}c^{2}}]
M_{W_{H}}.
\end{equation}

For the heavy gauge boson $X^{-}$, the main decay modes are
$\bar{q}q'$ and $l^{-}\nu$. The total decay width $\Gamma _{X^{-}}$
can be approximately written as[3]:
\begin{equation}
\Gamma
_{X^{-}}=\frac{\alpha_{e}M_{X}}{4S_{W}^{2}}(\delta_{t}^{2}+5\delta_{\nu}^{2}).
\end{equation}

In general, to reject backgrounds in hadronic collider environment,
the heavy gauge bosons may be most likely to be observed via their
pure leptonic decays. In this case, single production of the heavy
gauge boson $W_{H}^{-}$ or $X^{-}$ associated with a top quark gives
the possible observable five fermion final states with at least one
$b$ quark $\nu\nu l^{+}l^{-}b$ with $W_{H}^{-}(X^{-})\rightarrow
l^{-}\nu$ and $t\rightarrow W^{+}b\rightarrow \nu l^{+}b$. The
backgrounds of this kind signals mainly come from the $SM$ processes
$pp\rightarrow t\bar{t}+X$ and $pp\rightarrow tW^{-}+X$, in which
for the $t\bar{t}$ production process, one of the bottom quarks from
a top decay is assumed missing detection[7]. Under narrow width
approximation, the number of this kind signal events can be written
as $S_{W}=\pounds_{int}\sigma _{W}B r(W_{H}^{-}\rightarrow l^{-} \nu
)B r(t\rightarrow \nu l^{+}b)$ and $S_{X}=\pounds_{int}\sigma _{X}B
r (X^{-}\rightarrow l^{-}\nu)B r(t\rightarrow \nu l^{+}b)$, in which
$\pounds_{int}$ is the yearly integrated luminosity of the $LHC$
with $\sqrt{s}=14TeV$. In our numerical estimation, we will take
$\pounds_{int}=100fb^{-1}$.

\begin{figure}[htb] \vspace{-0.5cm}
\begin{center}
\epsfig{file=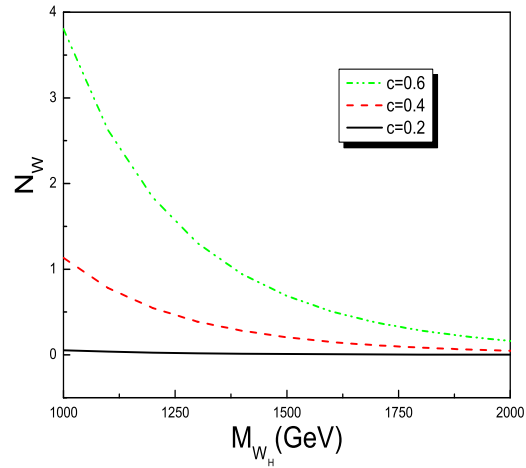,width=220pt,height=205pt}
\hspace{10cm}\vspace{-1.5cm}
 \caption{The ratio of signal over square root of the background ($S_{W}$/$\sqrt{B}$) for the
$W_{H}^{-}$ \hspace*{1.8cm} production associated with top quark at
the $LHC$. }
 \label{ee}
\end{center}
\end{figure}

To see the signals against the backgrounds for the heavy gauge
bosons $W_{H}^{-}$ and $X^{-}$, we introduce the ratio of signal
over square root of the backgrounds: $N_{W}=S_{W}/\sqrt{B}$ and
$N_{X}=S_{X}/\sqrt{B}$, in which $B$ includes the contributions of
$t\bar{t}$ production process and $tW^{-}$ production process. To
simple our calculation, we assume that all charged gauge bosons
$W_{H}^{-}$, $X^{-}$ and $W^{+}$ decay to $e\nu_{e}$. Similar to
Ref.[8], we take the appropriate cuts on the transverse momentum
$P_{T}(e)$ and rapidly $\eta(e)$ for the final electron as:
$P_{T}(e)>20GeV$ and $|\eta(e)|\leq2.5$. For the $b $ quark, which
comes from top quark decay, we take the cuts as: $P_{T}(b)>30GeV$
and $|\eta(b)|\leq 3$. Our numerical result are shown in Fig.4, in
which we plot the ratio $N_{W}$ as a function of the mass
$M_{W_{H}}$ for three values of the free parameter $c$. Our
numerical results show that the value of the ratio $N_{X}$ is
smaller than 0.1 in most of the parameter space of the $SU(3)$
simple model. So we do not give our numerical results for the heavy
gauge boson $X^{-}$ in Fig.4. One can see from Fig.4 that for $c\geq
0.5$ and $M_{W_{H}}\leq 1.5TeV$, the value of the ratio $N_{W}$ is
larger than 1, which might be detected at the $LHC[1]$.

Little Higgs models can be generally divided in two classes: product
group models and simple group models. All of the little Higgs models
predict the existence of the new charged gauge boson $W'$ with mass
in $TeV$ range, which has the $SM$-like couplings to the ordinary
fermions. In this letter, we investigate single production of these
new gauge bosons associated with a top quark at the $LHC$. We find
that the new charged gauge bosons $W_{H}^{-}$ and $X^{-}$, which are
predicted by the $LH$ model and the $SU(3)$ simple model,
respectively, can be abundantly produced at the $LHC$. However,
since the main backgrounds coming from the processes $pp\rightarrow
t\bar{t}+X$ and $pp\rightarrow tW^{-}+X$ are very large, the values
of the ratios $N_{W}$ and $N_{X}$ are very small in most of the
parameter space. The possible signals of the gauge bosons $X^{-}$
can not be observed via the process $pp\rightarrow gb+X\rightarrow
tX^{-}+X$ at the $LHC$. For the gauge boson $W_{H}^{-}$, it might be
possible to detect its signal via the process $pp\rightarrow
gb+X\rightarrow tW_{H}^{-}+X$ at the $LHC$ only for the mixing
parameter $c\geq 0.5$ and the mass parameter $M_{W_{H}}\leq 1.5TeV$.

\newpage

\noindent{\bf Acknowledgments}

C. X. Yue would like to thank the {\bf Abdus Salam } International
Centre for Theoretical Physics(ICTP) for partial support. This work
was supported in part by Program for New Century Excellent Talents
in University(NCET-04-0290), the National Natural Science Foundation
of China under the Grants No.10475037.

\null
\end{document}